\documentclass[aps,twocolumn,preprintnumbers]{revtex4}

\usepackage{subfigure}
\usepackage{multirow}
\usepackage{graphicx,xcolor}
\usepackage{fancyhdr}
\usepackage{longtable}
\usepackage{parskip}
\usepackage[T1]{fontenc}
\usepackage{dcolumn}   

\usepackage{bm}        
\usepackage{amsfonts}  
\usepackage{amsmath}   
\usepackage{amssymb}   

\usepackage{mathtools}

\DeclarePairedDelimiter\floor{\lfloor}{\rfloor}

\newcommand{\pwisein}{\left\{ \begin{array}{ll}}
\newcommand{\pwiseout}{\end{array}\right.}

\newcommand{\bp}{\boldsymbol{\epsilon}}
\begin{document}

\title{Robust photonic quantum gates with large number of waveguide segments}

\author{Khen Cohen, Haim Suchowski and Yaron Oz}

\affiliation{
School of Physics and Astronomy, Tel-Aviv University, Tel-Aviv 69978, Israel
}

\date{Jan 9, 2025}

\begin{abstract}
Realizing quantum information processors is challenged by errors and noise across all platforms. While composite segmentation schemes have been proposed in many systems, their application to photonic quantum gates in dual-rail configurations has only recently been demonstrated. However, prior research has been limited to a small number of segments, full noise correlation, and has overlooked the inherent power loss in such designs. Here, we study the fidelity and power loss of composite designs for photonic quantum gates with a high number of segments of varying geometrical widths. Using numerical simulations, we analyze the relationship between gate performance and the number of waveguide segments, accounting for statistical error correlations and variances.
Beyond effectively reducing the errors, an asymptotic scaling pattern of quantum gate fidelity and power loss is observed as the number of segments increases. This analysis is examined in Silicon and Lithium Niobate platforms, addressing practical implementation challenges. Our findings demonstrate that optimized multi-segment waveguide geometrical designs significantly enhance the robustness and efficiency of photonic quantum gates, paving the way for more reliable quantum information processors.


\end{abstract}

\maketitle

\section{Introduction}

The realization of a practical quantum information processor remains a great challenge, as errors and noise inevitably arise during the implementation of its core elements. These challenges are universal across various physical platforms for quantum information processing \cite{bennett2000quantum, Nielsen2000}, including superconducting circuits \cite{Kjaergaard2020}, trapped ions \cite{Bruzewicz2019}, and photonic systems \cite{Flamini2018, giordani2023integrated}. Among these, integrated photonics, particularly dual-rail waveguide platforms, has emerged as a promising approach due to its scalability and compatibility with existing fabrication technologies.

In dual-rail photonic systems, errors primarily stem from fabrication imperfections \cite{Kyoseva2019}, inaccurate coupling strengths, and environmental interactions \cite{noiseoneoverf}, which compromise the fidelity of the quantum state preparation and quantum gates, the fundamental building blocks of any quantum algorithm \cite{Nielsen2000}.
For example, deviations in waveguide geometries from the intended design, such as variations in waveguide width, can significantly impact the performance of dual-rail single-photon and two-photon gates. These deviations induce Pauli errors, which in turn lead to a reduction in gate fidelity \cite{melati2014real}.
Various strategies have been developed to address these challenges \cite{QNoiseMitigation1, NoiseMitigation2}. Error correction codes, for example, construct logical qubits from entangled physical qubits to reduce errors \cite{QEC1}.
However, these methods impose stringent fidelity thresholds on the underlying physical gates, requiring a fidelity error of less than $1\%$ in each of the physical photonic gates \cite{NoiseThreholdQC, QuantumErrorCorrection, shi2022high, noiri2022fast}.

One promising error mitigation approach, which reduces physical errors through advanced design and implementation techniques, is the composite design method, originally developed in atomic physics and nuclear magnetic resonance (NMR) \cite{CPNMR, CPNMR2, CPAtomicPhysics, CPAtomicPhysics2}. This method employs systematic parameter variations to counteract errors during system evolution. In NMR, large number of pulses were suggested, using symmetric and antisymmetric composite pulses designs \cite{HUSAIN2013145}, and a smooth composite pulses scheme have been proposed in Ref. \cite{SmoothCP}.

Recent work has adapted this approach to integrated photonics, demonstrating the potential for high-fidelity dual-rail quantum gates by segmenting waveguides and varying their geometries \cite{Kyoseva2019, Kaplan2023}. However, prior studies primarily considered designs with a small number of segments and assumed fully correlated fabrication errors \cite{Kaplan2023}. Subsequent findings indicate that effective error mitigation requires the correlation strength to remain below a specific threshold \cite{YaronIdoNoiseThreshold}. Moreover, earlier analyses did not evaluate the power loss associated with segmentation, an essential factor for practical implementation.

\begin{figure*}
    \centering
    \includegraphics[width=1.0\linewidth]{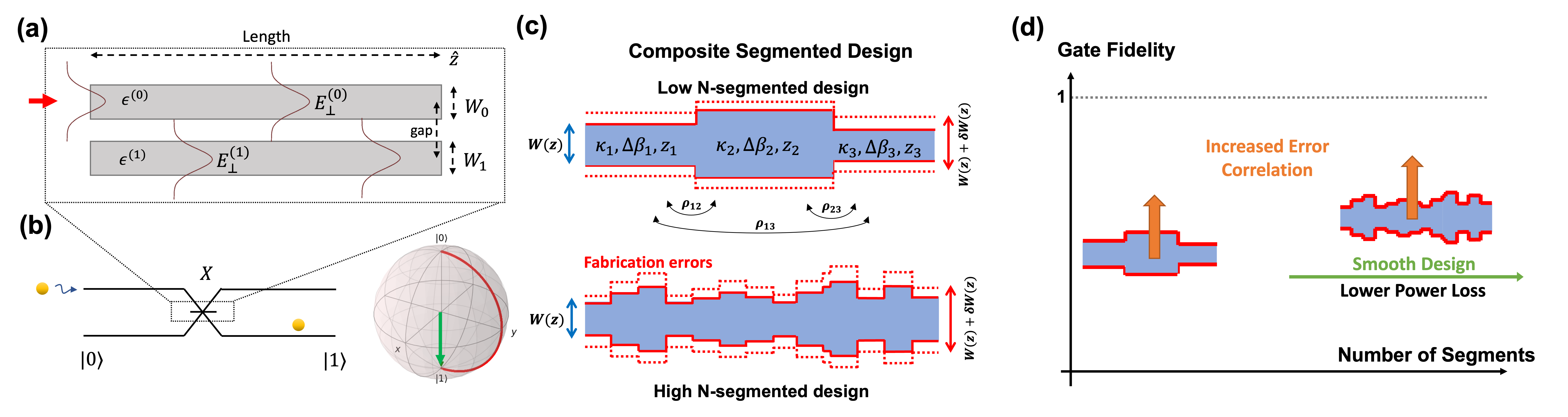}
    \caption{Comparison of composite segment design with variable number of segments. (a) Two coupled waveguides, with widths denoted as $W_0$ and $W_1$, a gap between them and a defined length. (b) Optical waveguide gates in dual-rail encoding for $X$ (Pauli), gate, along with its graphical representation on the Bloch sphere. (c) Fabrication noise model with spatial correlation between segments, for a small and large number of segments. (d) Comparison of Fidelity and Power Loss between a small and large number of segments. Increasing either the number of segments or the noise spatial correlation improves Fidelity. Our proposed smooth design significantly reduces power loss, in addition to increasing Fidelity.}
    \label{fig:main_diag}
\end{figure*}

Here, we study the fidelity and power loss of photonic quantum gates composed of increased number of waveguides' segmentations. 
We find that as the number of segments increases, the errors associated with variations in waveguide geometrical widths are significantly mitigated, leading to substantially higher gate fidelity.
Using numerical simulations on Silicon-on-Insulator (SOI) and Thin-Film Lithium Niobate (TFLN) platforms, we demonstrate that increasing the number of waveguide segments $N$, not only improves the average quantum gate fidelity $\mathbf{E}\left[F\right]$ but also reduces its standard deviation $\sigma_F\left[ F \right]$, with a specific scaling with $N$.

Furthermore, we analyze the impact of segmentation on power loss due to back-scattering effects. Our analytical and numerical results show that the power loss increases linearly with the number of segments, underscoring the necessity of employing smoothing techniques for designs with large $N$. To address this, we propose smooth segmented photonic quantum gate designs based on adiabatic and Fresnel reflection principles. These designs effectively reduce power loss while maintaining high fidelity performance. Finally, we conclude that a segmentation number $N \sim O(10)$ is sufficient for most practical implementations, providing a balance between fidelity enhancement and manageable power loss. This work advances the development of practical quantum dual-rail gates for photonic integrated platforms, and quantum information processors in integrated photonics.

The paper is organized as follows: Section II introduces the theoretical background, including dual-rail encoding within the framework of integrated optics, as well as the necessary background on composite schemes. In Section III, we formulate the optimization problem and define the cost function, incorporating the optimization range, quantum gate fidelity, and smooth design conditions aimed at minimizing power loss. Section IV presents the optimization results for gate fidelity and power loss across varying numbers of segments and different smooth design configurations. Finally, Section V provides a summary and outlook. Additional technical details are included in four appendices.

\section{Theoretical background}
\label{sec:theoretical_bg}

\subsection{Dual-rail realization for of two-mode dynamics}
\label{subsec:dual_rail}

In the integrated dual-rail realization, two closely spaced optical waveguides form a coupled system that emulates the dynamics of a quantum two-level system \cite{AmnonYariv}. This system, governed by coupled-mode theory (CMT), is described by a unitary propagator that captures its evolution over the propagation length:
\begin{gather}
    i \partial_z\left(\begin{array}{l}
E_0(z) \\
E_1(z)
\end{array}\right)=\left(\begin{array}{cc}
-\Delta(z) & \kappa^*(z) \\
\kappa(z) & \Delta(z)
\end{array}\right)\left(\begin{array}{l}
E_0(z) \\
E_1(z) 
\end{array}\right) \ ,
\end{gather}
where $E_0(z)$ and $E_1(z)$ are the electrical fields at the of being in states $|0\rangle$ and $|1\rangle$, the mode mismatch $\Delta (z)$ arises from differences in the effective refractive indices, $n_{\text{eff},1}$ and $n_{\text{eff},2}$ of the waveguides: 
\begin{eqnarray}
    \Delta (z) = \frac{\beta_1 (z)-\beta_2(z)}{2} \equiv \frac{2\pi}{\lambda}\left( n_{{eff}_1}(z) - n_{{eff}_2}(z) \right) \ ,
    \label{cw2}
\end{eqnarray}
where $\beta_1$ and $\beta_2$ are the propagation constants of the two waveguides.

The coupling coefficient $\kappa$, resulting from the evanescent field overlap between the waveguide modes, is defined as:
\begin{eqnarray} \label{eq:kappa_calc}
    \kappa (z)= \frac{c}{4 \pi \lambda} \iint &  \left[\epsilon(x, y, z)-\epsilon^{(1)}(x, y, z)\right]\cdot  &  \\ \nonumber
    & \vec{E}_{\perp}^{(0)}(x,y,z)\vec{E}_{\perp}^{(1)}(x,y,z) \ d x d y  & \ ,
\end{eqnarray}
where $(0)$ and $(1)$ refer to the two waveguides, $(x,y)$ are the transverse directions, $z$ denotes the propagation direction, $\vec{E}_{\perp}^{(0)}(x,y,z)$ and $\vec{E}_{\perp}^{(1)}(x,y,z)$ are the transverse electric field modes for each of the waveguides, and $\epsilon(x,y, z)$  and $\epsilon^{(1)}(x,y, z)$ are the permittivity distributions of the coupled waveguides system (namely, the profile of the two waveguides together) and of waveguide $(1)$ alone, respectively \cite{AmnonYariv}. $c$ is the speed of light and $\lambda$ is the wavelength. A diagram of two coupled waveguides is shown in Fig. \ref{fig:main_diag} a.

The unitary propagator of such a system is:
\begin{gather}
    U(z, 0)=\mathcal{Z}\left[\exp \left[-i \int_0^z\left(\begin{array}{cc}
-\Delta\left(z^{\prime}\right) & \kappa^*\left(z^{\prime}\right) \\
\kappa\left(z^{\prime}\right) & \Delta\left(z^{\prime}\right)
\end{array}\right)\right] d z^{\prime}\right] \ ,
\end{gather}
where $\mathcal{Z}$ denotes z-ordering. In cases where $\kappa$ and $\Delta$ are independent of z, the propagator simplifies to:
\begin{gather}
    U(z, 0)=\exp \left[-i z\left(\begin{array}{cc}
-\Delta & \kappa^* \\
\kappa & \Delta
\end{array}\right)\right]
\end{gather}
Alternatively, using the well-known relation $e^{-i\alpha \vec{\sigma}\cdot \hat{n}}=cos(\alpha)\mathbf{I}+\vec{\sigma}\cdot \hat{n} sin(\alpha)$, for a unit vector $\hat{n}$, and $\vec{\sigma}=(\mathbf{X},\mathbf{Y},\mathbf{Z})$ a vector of Pauli matrices, the unitary operator can be written as:
\begin{gather}
    \label{eq:constant_wg}
 U\left(\kappa, \Delta, z\right)= \cos \left(\Omega_g z\right) \mathbf{I}- 
i \frac{\kappa}{\Omega_g} \sin \left(\Omega_g z\right) \mathbf{X} \nonumber\\
+ i \frac{\Delta}{\Omega_g} \sin \left(\Omega_g z\right) \mathbf{Z} \ ,
\end{gather}
while we use use the transfer frequency determined by the effective coupling parameter $\Omega_g = \sqrt{\kappa^2 + \Delta^2}$.

The dual-rail encoding in two coupled optical waveguides forms a quantum two-level system, where light propagation emulates $SU(2)$ dynamics, and is elegantly represented on the Bloch sphere (Fig. \ref{fig:main_diag} (b)), where the evolution of light in the coupled waveguides mirrors Rabi oscillations in atomic physics. For symmetric configurations ($\Delta = 0$), the system achieves Complete Power Transfer (CPT), between the two levels. This $SU(2)$ framework underpins the implementation of integrated photonic quantum gates and sets the stage for advanced multi-qubit operations via concatenated or segmented couplers.

Realizing two-qubit gates, such as the Controlled-NOT (CNOT) or Controlled-Z (CZ) gates, typically requires additional ancilla modes, measurements, and linear optical elements, and these gates often exhibit a non-deterministic nature \cite{KLM}. A well-known example is the CNOT gate, which can be implemented using beam splitters, phase shifters, and single-photon detectors \cite{Ralph2002}.

By carefully arranging these components, conditional operations are performed, where the state of one qubit (the control qubit) governs the state of another (the target qubit). The universality of quantum computing in the dual-rail encoding scheme is achieved through the combination of single-qubit and two-qubit gates. This universal gate set enables the construction of arbitrary quantum circuits, paving the way for the implementation of complex quantum algorithms.

\subsection{Coherent Noise Analysis}

\begin{figure}
    \centering
    \includegraphics[width=1.0\linewidth]{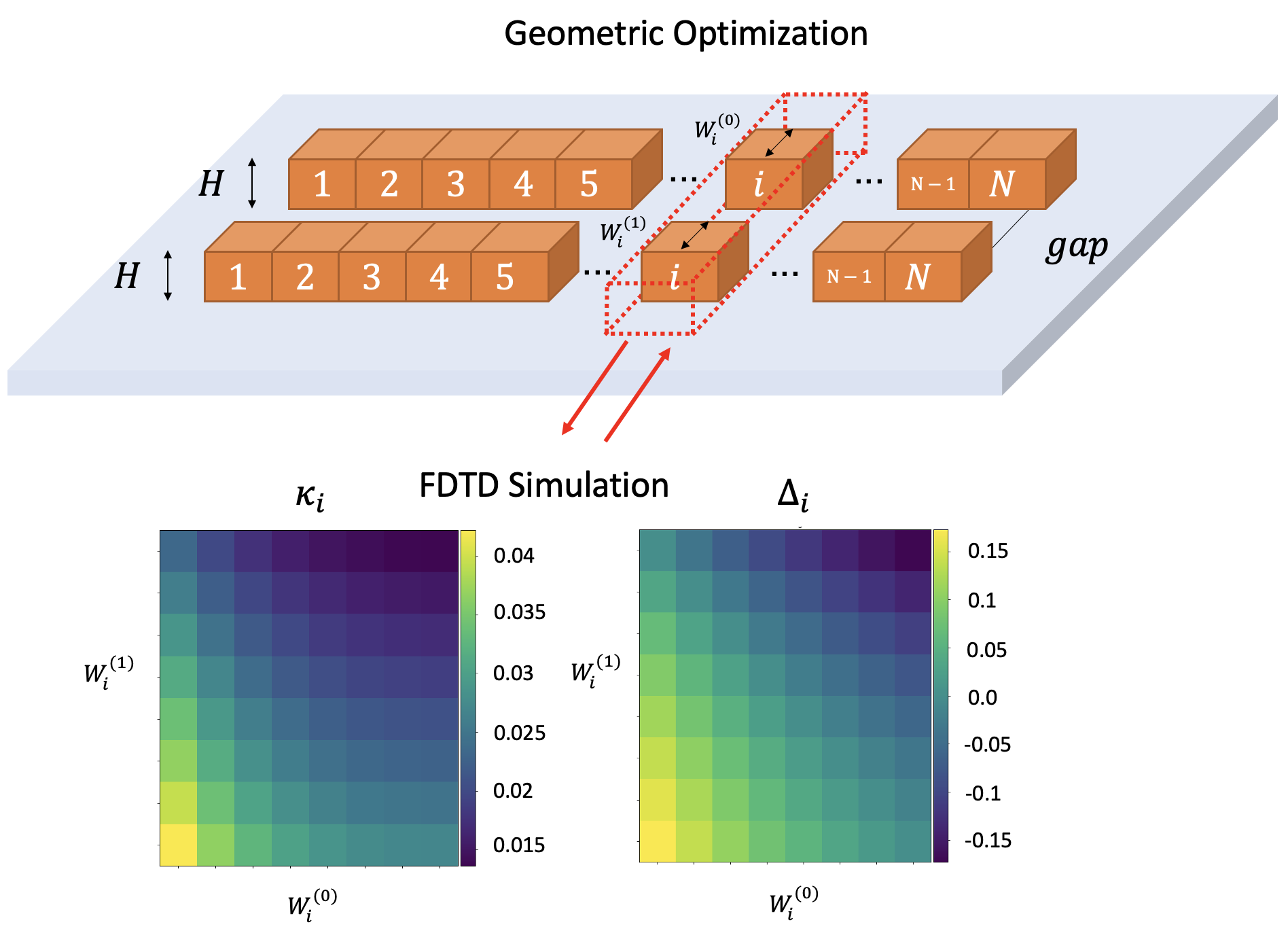}
    \caption{$N$ segments composite configuration. We divide each of the two waveguides to $N$ segments. Each segment $i$ is characterized by its coupling $\kappa_i$, mode-mismatch coefficient $\Delta_i$ and length $z_i$. The interpolation map for $\kappa$ and $\Delta$ as a function of the waveguides widths $W_0, W_1$, in $nm$ units. These results are extracted from a Lumerical FDTD Eigensolver simulation, with geometry resolution of $5 nm$, and a bilinear interpolation to evaluate the values in-between.}
    \label{fig:small_segments_wg}
\end{figure}

Coherent noise in photonic quantum gates, caused by fabrication imperfections, introduces systematic errors in the form of random unitary rotations. In the photonic regime, these errors are primarily attributed to variations in the waveguide widths during the fabrication process, which significantly impact the coupling coefficient $\kappa$ and mode mismatch $\Delta$. Additionally, due to the nature of errors arising during the fabrication of coupled waveguides, these errors are fully correlated between the two coupled waveguides, further compounding their effect on gate fidelity \cite{ExperimentCP}. Such systematic deviations degrade gate performance and accumulate in multi-qubit operations, necessitating mitigation to meet the stringent thresholds required for quantum error correction and measurement-based quantum computation \cite{Nielsen2000}.

To quantify the extent to which the noisy gate deviates from the ideal gate, we define the gate fidelity $F$ by the trace norm:
\begin{eqnarray}
F\left(U_{\text{ideal}}, U(\boldsymbol{\epsilon})\right) = \frac{1}{2} \left|\text{Tr}\left(U_{\text{ideal}}^\dagger U(\boldsymbol{\epsilon})\right)\right|  \ ,
\end{eqnarray}
where $U_{\text{ideal}}$ represents the desired ideal unitary gate, and $U(\boldsymbol{\epsilon})$ represents its physical realization, which depends on a set of $m$ jointly distributed random errors denoted as $\boldsymbol{\epsilon} = \{\epsilon_a\}_{a=1}^m$. The fidelity is a random function that takes values in the interval range $[0, 1]$, where zero and one correspond to the absence of errors and to the maximal deviation from the desired quantum gate operation, respectively.

The primary objective of our analysis in the context of error mitigation, is to maximize the mean fidelity, denoted as $\mathbf{E}\left[F\right]$, and to minimize its standard deviation, denoted as $\sigma[F]$:
\begin{eqnarray}
\mathbf{E}\left[F\right] &=& \mathbb{E}_{\bp}\left[F\left(U_{\text{ideal}}, U(\boldsymbol{\epsilon})\right)\right] \ ,
\nonumber\\
\sigma[F] &=& \sqrt{\mathbb{E}_{\bp}\left[F\left(U_{\text{ideal}}, U(\boldsymbol{\epsilon})\right)^2\right] - \mathbf{E}\left[F\right]^2} \ .
\end{eqnarray}

Implementation of the composite pulse error mitigation scheme in integrated photonics, or in any other physical platform, is accompanied by power loss due to several reasons such as absorption or scattering. The back-scattering is a direct consequence of the non-smooth profile of the segmented waveguides width profiles,
and captures the total power loss (see Appendix \ref{app:power_loss_sim_comp}).
The gate reflection power loss is total power transmission coefficient $P_t$:
\begin{equation} \label{eq:power_transmission}
    P_{t} \equiv T = \prod_{k=1}^{N-1} T_k (\kappa_k, \Delta_k, z_k, \epsilon_k) \ ,
\end{equation}
where $T_k$ is the power transmission coefficient from the segment $k$ to the segment $k+1$.

Due to the changes in the effective dispersion coefficients between the segments of the waveguides, we estimate the back-scattering by the Fresnel equation, with effective refractive indices of the two consecutive segments:
\begin{gather}
    T_k = 1- \left( \frac{n_{\text{eff},k} - n_{\text{eff},k+1}}{n_{\text{eff},k} + n_{\text{eff},k+1} } \right)^2 \ ,
\end{gather}

The total reflection coefficient is given by:
\begin{gather}
    \label{R}
    R = 1 -P_t \ .
\end{gather}

\subsection{The Composite Segmentation Method}
The composite pulses method is well-known for its ability to mitigate errors by decomposing a single gate into a sequence of smaller unitary operations. By strategically leveraging the correlation between error factors, this sequence can be meticulously designed to cancel out errors effectively. In the dual-rail implementation, as demonstrated in Ref. \cite{Kaplan2023}, a two-waveguide coupler is divided into multiple sections with varying widths and shorter propagation lengths.

We define the gate $U(\bp)$ as a sequence of $N$ segments:
\begin{equation}
U_{cp}(\vec{\bp})= \prod_{i=1}^N U_i(\bp_i) \ ,
\label{CP}
\end{equation}
Where $\bp_i$ are the error parameters (in our context, in the width of the waveguides), a single one for each gate. The $U_i$ are characterized by a its own $\kappa_i, \Delta_i, z_i$ parameters (coupling coefficient, mode mismatch, and length, respectively).

Therefore, designing a robust unitary gate $U_{cp}$ is actually finding the set of geometrical parameters (width, length, refractive index etc.), that determines $\{\kappa_i, \Delta_i, z_i\}_{i=1}^3$ (total of $3N$ degrees of freedom) such that the gate will be more robust to the presence of noise, namely, the fidelity between $U_{cp}$ and $U$ will be as close as possible to 1.

\section{High fidelity quantum gates using Large N segmentation}

In this study, we focus on dual-rail waveguide platforms, presenting a detailed study of segmented and smooth waveguide designs of robust, high-performance quantum gates designs. Fig. \ref{fig:main_diag} provides an overview of the work, including the key results from this paper: showcasing the relationship between fidelity and power loss against the number of segments (see Fig. 1c), as well as the impact of smoothing and noise correlation (see Fig. 1d).

Our aim is to obtain high-fidelity quantum gates with as smooth as possible design that will minimize the power loss, for any arbitrary number of segments $N$.
As discussed, abrupt changes in the waveguides widths increase the power loss due to back scattering, and also
give rise to higher spatial modes in the waveguide that distort the signal, hence changing the qubit state. Moreover, waveguide designs must be physically implementable subject to the manufacturing resolution. 

We divide each waveguide into a sequence of $N$ small segments, each of length $z_i$ as shown in Fig. \ref{fig:small_segments_wg}. The two waveguides are separated by a gap, defined as the distance between the centers of the waveguides, and have a height $h$.
The optimization process involves three stages. Firstly, we construct a look-up table with the Lumerical FDTD simulator to map geometries to the coupling coefficient and mode-mismatch parameters inversely. Next, we establish the optimization cost function, including regularizes. Finally, we filter-out devices that are not physically realizable. The subsequent subsections will cover these stages in depth.

\subsection{The Fidelity Term}
As discussed, in our analysis we assume that the errors in the two waveguides widths $W_0$ and $W_1$ 
are fully correlated, and can be sampled from a Normal distribution with zero mean and a standard deviation $\sigma$, $D\sim\mathcal{N}(0,\sigma)$. For our analysis, we consider the dimensionless parameter $\sigma_\% \equiv 100 \frac{\sigma}{W_{mean}}$, where
\begin{equation}
W_{mean}\equiv \frac{1}{2N}\sum_i \left( W_{i}^{(0)}+W_{i}^{(1)} \right) \ ,
\end{equation}
and we focus on the range $1\%-10\%$, which is approximately $10_{nm}-50_{nm}$.
The fidelity loss term reads:
\begin{equation}
    \mathcal{L}_F = 1 - E_{\delta W\sim D}\Big[ F\left(U_{\text{ideal}}, U \right) \Big] \ ,
    \label{F}
\end{equation}
where $U_{ideal}$ is the ideal quantum gate and $U$ is its physical realization.

\subsection{The Range of Width and Length}

We add a regularization in order to restrict the width of the waveguides to be in a range between the minimum and the maximum of an allowed region:
\begin{eqnarray}
    \mathcal{L}_{\Omega} = \lambda_W \sum_i &\bigg(& \max{ \left(W_i-W_{max}, 0\right)^2} + \nonumber 
    \\ &+& \min{ \left(W_{min}-W_i, 0\right)^2})\ \bigg) \ ,
    \label{Omega}
    \end{eqnarray}
where $\lambda_W$ is a hyperparameter for the minimal and maximal widths.
We define the optimization domain $\Omega=(\Omega_{dz}, \Omega_w)$, as the set of all experimentally realizable waveguide widths. The number $N$ of segments in the design varies between 1 and 200, and we denote by $L$ the total length of the design. The degrees of freedom for the optimization are the set of widths and lengths of each segment: $\left\{W^{\text{L}}_{i}, W^{\text{R}}_{i}, dz_{i} \right\}_{1\leq i \leq N}$.

The waveguide width initialization is sampled uniformly in the range between the maximum and minimum of the
feasible domain. Furthermore, strong regularization is activated once the design is out of the optimization domain $\Omega$.

To avoid overly short segments that could violate our assumptions and necessitate high fabrication resolution, we restricted our focus to devices with sufficiently large segment lengths. Specifically, we ensured that the dimensionless parameter characterizing propagation, defined as the ratio of each segment's length to the wavelength in the material, remains sufficiently large.

While it is also possible to impose an additional regularization on the total design length $L$, this restriction is not applied in our analysis. Instead, filtered-out devices with a minimum segment length less than $1 \mu m$.
Our results show that nearly all designs naturally exhibit total lengths in the range of $10 \mu m$ to $100 \mu m$ (see the table of our designs \ref{tab:result_table} in the Appendix).

\subsection{Smooth Design Conditions}

We define the \textit{baseline optimization}, as a search for the highest fidelity designs, without taking into account any smoothness requirement. This lead to designs with high back-scattering, and power loss.
The alternative optimized designs will be evaluated in comparison to these baseline configurations.

\subsubsection{Minimizing Fresnel Reflection}
In order to reduce the power loss, we add a regularization term, which reduces the back reflection coefficient. Thus, among the robust high-fidelity design parameter space, we search for the designs with the minimal power loss. The total cost function reads: 
\begin{equation}
    \mathcal{L} =   \mathcal{L}_F +  L_{\Omega} +  \lambda_R R \ ,
\end{equation}
where  $\mathcal{L}_F, L_{\Omega}, R$ are given by Eq. \ref{F}, \ref{Omega} and \ref{R}, respectively.

\subsubsection{Imposing An Adiabaticity Condition}
The adiabatic theorem guarantees, under certain conditions,  a smooth transition between the ground states of two different Hamiltonians \cite{adiabatic}. In our setup we impose an adiabatic condition on the 
design in the form:
\begin{equation}
    \left|\frac{\partial \beta }{\partial z}\right| \ll \beta(z) \ .
\end{equation}
Assuming a power law dependence of $\beta$ on the cross-section area:  $\beta = a W^m$ for some $a$ and $m$, where $W$ is the width of the waveguide, we get:
\begin{equation}
    \left|\frac{\partial W / \partial z} {W(z)}\right| \ll 1  \ .
\end{equation}
We define a corresponding regularization term:
\begin{equation}
    A = \max_i \frac{ \left|W_{i+1} - W_i\right|} {\left(\frac{W_{i+1} + W_i}{2}\right)\left(\frac{z_{i+1} + z_i}{2}\right)} \ .
\end{equation}
The cost function now reads:
\begin{equation}
    \mathcal{L} =  \mathcal{L}_F +  L_{\Omega} +  \lambda_A A \ .
\end{equation}

\subsection{Segments Error Correlation Model}
We also account for cases with partial correlation, where errors in different segments may exhibit varying degrees of dependency. To assess the impact of correlation strength on design performance, we introduce an error correlation matrix to model the relationships between errors across $ k,l \in \{1, ... N \}$ segments:
\begin{equation}
    \rho_{kl}(\mu) = \begin{cases}
        1 & k = l \\
        \frac{1}{\left(\left|k-l\right| + 1/2 \right)^{2\mu} + 1 } & k \neq l \ .
    \end{cases}
    \label{rho}
\end{equation}
The dimensionless parameter $\mu$ controls the strength of the correlation: $\mu=0$ implies strong correlations between segments while $\mu \rightarrow \infty$ implies no correlation. The results of our correlations simulation will be discussed in the simulation section (see sub-section \ref{subsec:fidelity_and_error_correlation}).

We optimize the design for two materials: Silicon and $\text{LiNbO}_{3}$ (Lithium Niobate). Silicon, widely used in photonic applications, is notable for its high refractive index of approximately 3.48. $\text{LiNbO}_{3}$, on the other hand, is a promising candidate due to its exceptional optical properties, including a wide transparency window ($0.35 \mu m$ to $5 \mu m$), large optical nonlinearity, high electro-optic coefficient, and strong acousto-optic effect. We assume in our analysis that only the fundamental $TE_{01}$ mode is occupied.

For both materials we use a gap of $1.0~ \mu m$ and the width of each waveguide is varied between $500~ nm$ to $850~ nm$. We further assume the manufacture resolution (smallest resolution in the optimization) to be $20~ nm$, waveguide height of $220~ nm$, etch of $150~nm$ and wavelength of $1.55~\mu m$. To evaluate the mode-mismatch and coupling coefficients of the design we use Lumerical Eigensolver simulation (as shown in Fig. 
\ref{fig:small_segments_wg}).

\section{Optimization Results}

In the optimization we divide each waveguide into $N=100$ segments. We use the PyTorch Package, with Adam optimizer, CosineAnnealing LR scheduler, with an initial learning rate of 0.1, batch size of 1024 samples, and the number of epochs was set to be around 5000. The length of the device is randomly initialized between $25~\mu m$ and $100~\mu m$. In our analysis we focus on the Pauli $X$ (without mode-mismatch) and the Hadamard $H$ (with mode-mismatch) single-qubit quantum gates. 

In this study, we observed that the scaling behavior of the two materials is quite similar, as well as the gate types. Consequently, we decided to concentrate on presenting our analysis for only one material, and one gate at a time.

To visually represent the geometric design achieved through the optimization process, refer to Fig. \ref{fig:power_dist_prof}. For a depiction of the trajectories on the Bloch sphere, please see Appendix \ref{app:bloch_sphere_trajectories}.

\subsection{Mean Fidelity and Variance}

\begin{figure}
    \centering
    \includegraphics[width=1.0\linewidth]{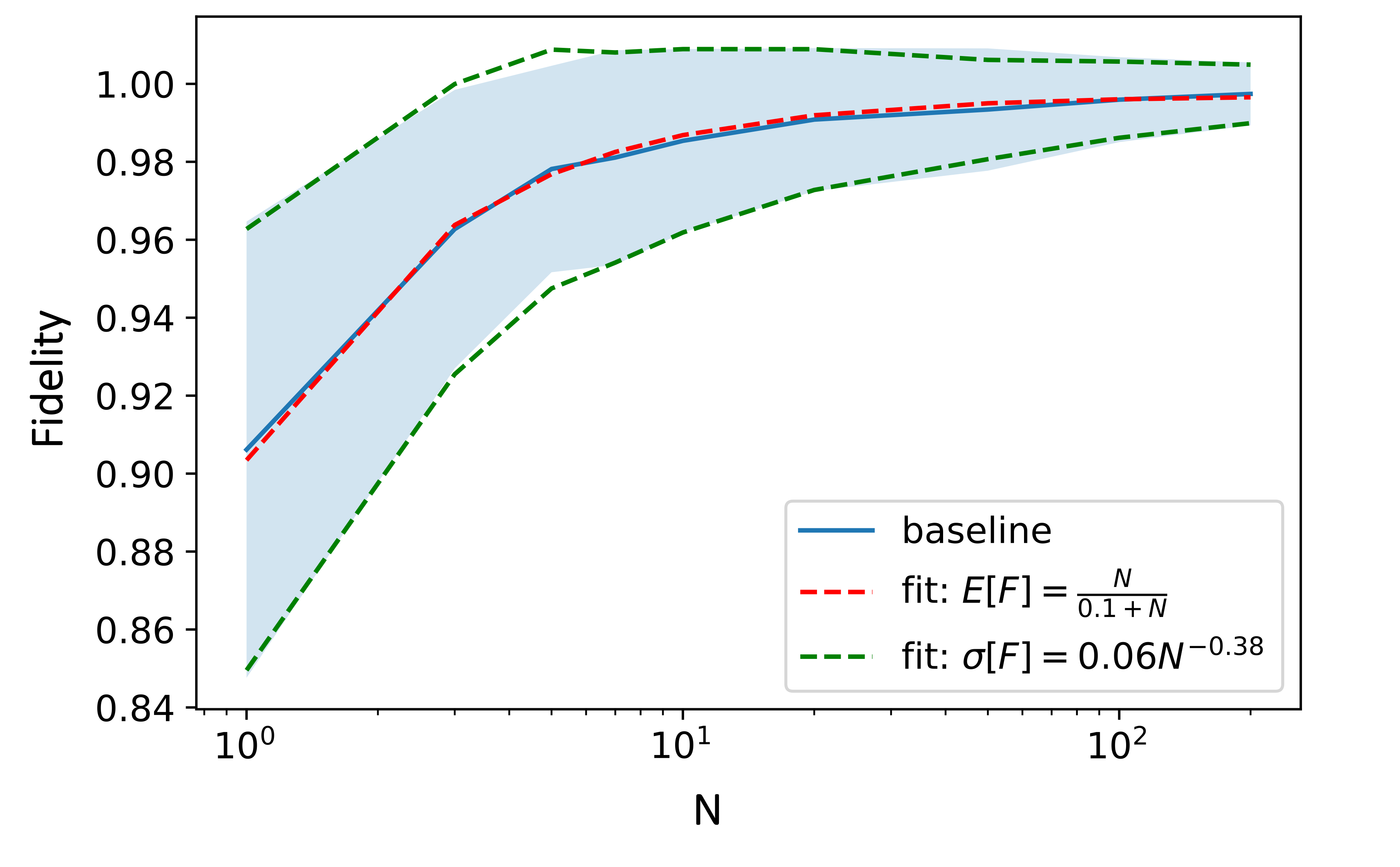}
    \caption{Mean Fidelity and standard deviation (shaded area) versus the number of segments for $\text{LiNbO}_{3}$ X gate, assuming Normally distributed width noise with $\sigma=30_{nm}$ ($\sigma_\% \approx 5\%$). We see that the mean fidelity increases and variance decreases as the number of segments grows. This plot was created using an ensemble average over 20 optimizations for each $N$.}
    \label{fig:FvsN}
\end{figure}

\begin{figure}
    \centering\includegraphics[width=1.0\linewidth]{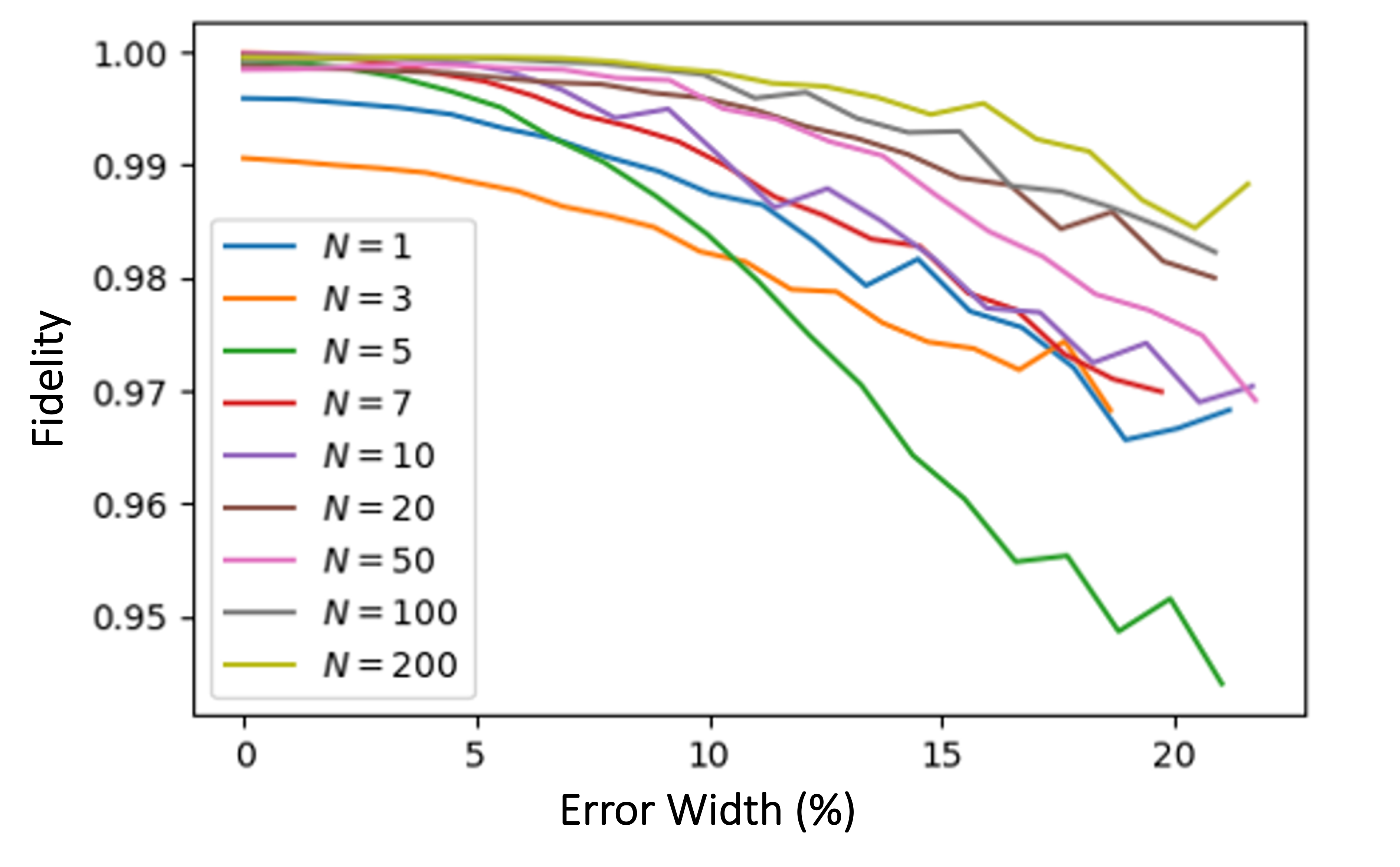}
    \caption{The expectation value of the fidelity for different number of segments, for increasing noise width, in $\text{LiNbO}_{3}$ X gate. The design is optimized for $\sigma_\% \approx 5\%$. Increasing the number of segments leads to a higher robustness for large width errors.}
    \label{fig:FvsE}
\end{figure}

Fig. \ref{fig:FvsN} shows the expectation value and standard deviation of the fidelity for different numbers of segments taken in the range 1 to 200. The presented value for each segment is an average over 20 optimizations values.
Increasing the number of segments leads to
an increase in the gate fidelity. The most significant improvement is obtained when the increasing the number of segments from 1 to 10. In Fig. \ref{fig:FvsE} we show the expectation value of the fidelity for different error magnitudes. Empirically, we find that the fidelity for Adiabatic and Fresnel designs are very similar to the baseline.
We numerically observe large $N$ scaling laws for the expectation value and variance of the fidelity as functions of the number of segments:
\begin{gather}
    \mathbf{E}\left[F\right] \sim 1 - \frac{a}{N},~~~~\sigma[F] \sim N^{-b} \ ,
\end{gather}
for some constants $a$ and $b$, which we expect to be non-universal and depend on the geometry of the device, the material and the wavelength. 
In our example, they are $a=0.1$, and $b=0.38$.

\subsection{Power Loss}

To evaluate the power loss of the design, we use Lumerical simulator to measure the transmitted light for each transition between two neighboring segments with different widths. In appendix \ref{app:power_loss_sim_comp} we present a comparison between different power loss calculation methods (Fresnel coefficients, FDTD, varFDTD and EME simulations), while here we focus on the comparison between the Fresnel coefficient (up to a global factor, see appendix \ref{app:power_loss_sim_comp} for details), and the FDTD simulation result, which is considered the most accurate among the simulation methods. 

\begin{figure}
    \centering
    \includegraphics[width=1.0\linewidth]{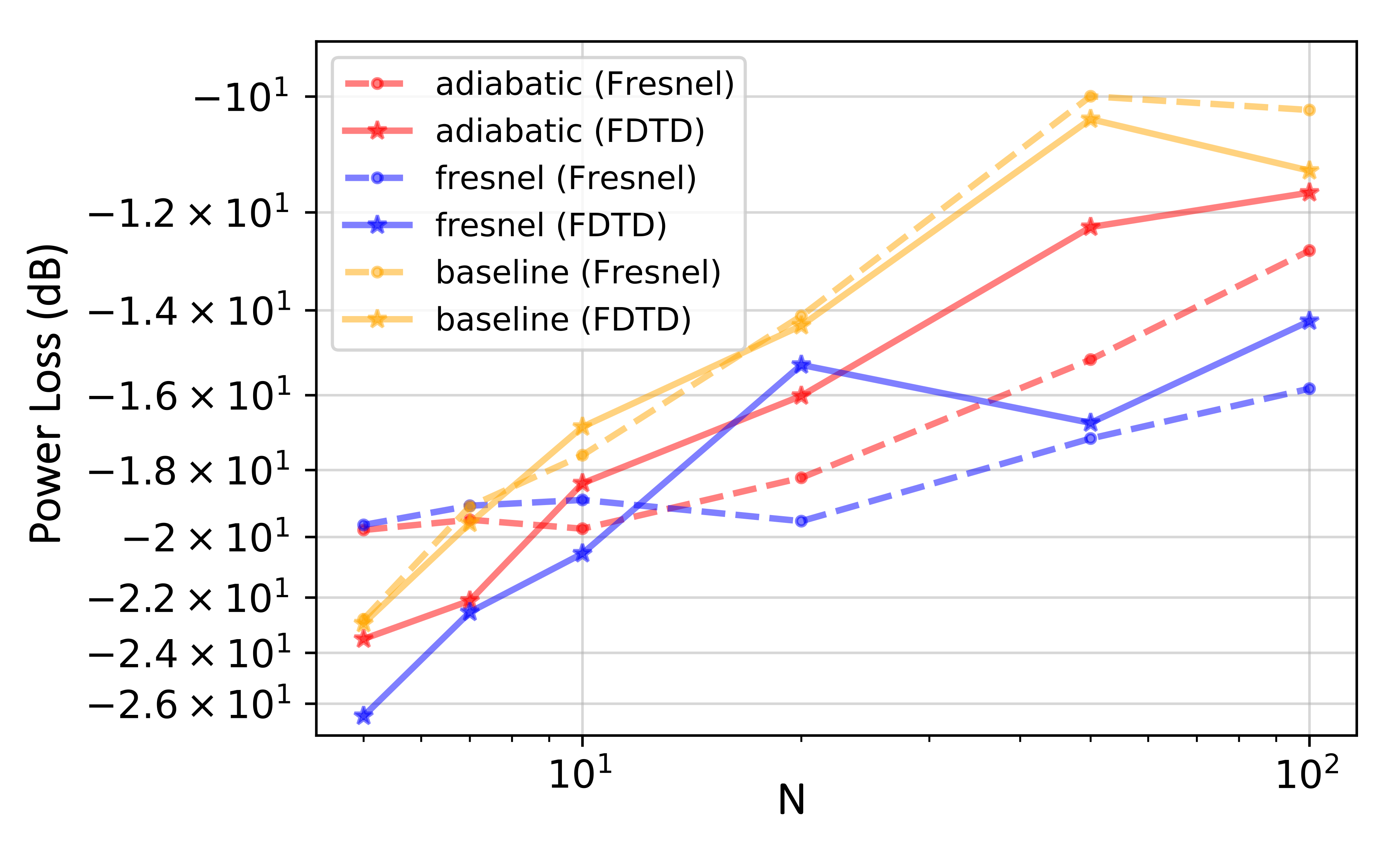}
    \caption{Power loss versus number of segments for Hadamard, in $\text{LiNbO}_{3}$ based Hadamard-Gate, using Fresnel coefficients (up to a global scale factor) in dashed line and FDTD simulation in solid line. We see an increase in the power loss as the number of segments increases, with an improvement when employing the Fresnel and the Adiabatic smoothing methods. This plot is created using an ensemble average over 10 optimizations for each $N$ and optimized for $\sigma_\% \approx 5\%$.}
    \label{fig:PvsN}
\end{figure}

Fig. \ref{fig:PvsN} shows the power loss for different number of segments, for the Hadamard gate made of $\text{LiNbO}_{3}$, each data point is averaged over 10 optimizations.. While there are some differences between the Fresnel coefficients and the FDTD, their overall behavior is very similar. We find empirically that optimizing over the Fresnel coefficients leads to a reduction in the total power loss estimated by the FDTD.
To calculate the total power loss per design, we converted the loss per transition (between segments) using Eq. \ref{eq:power_transmission}.

\begin{figure*}
    \centering
    \includegraphics[width=1.0\linewidth]{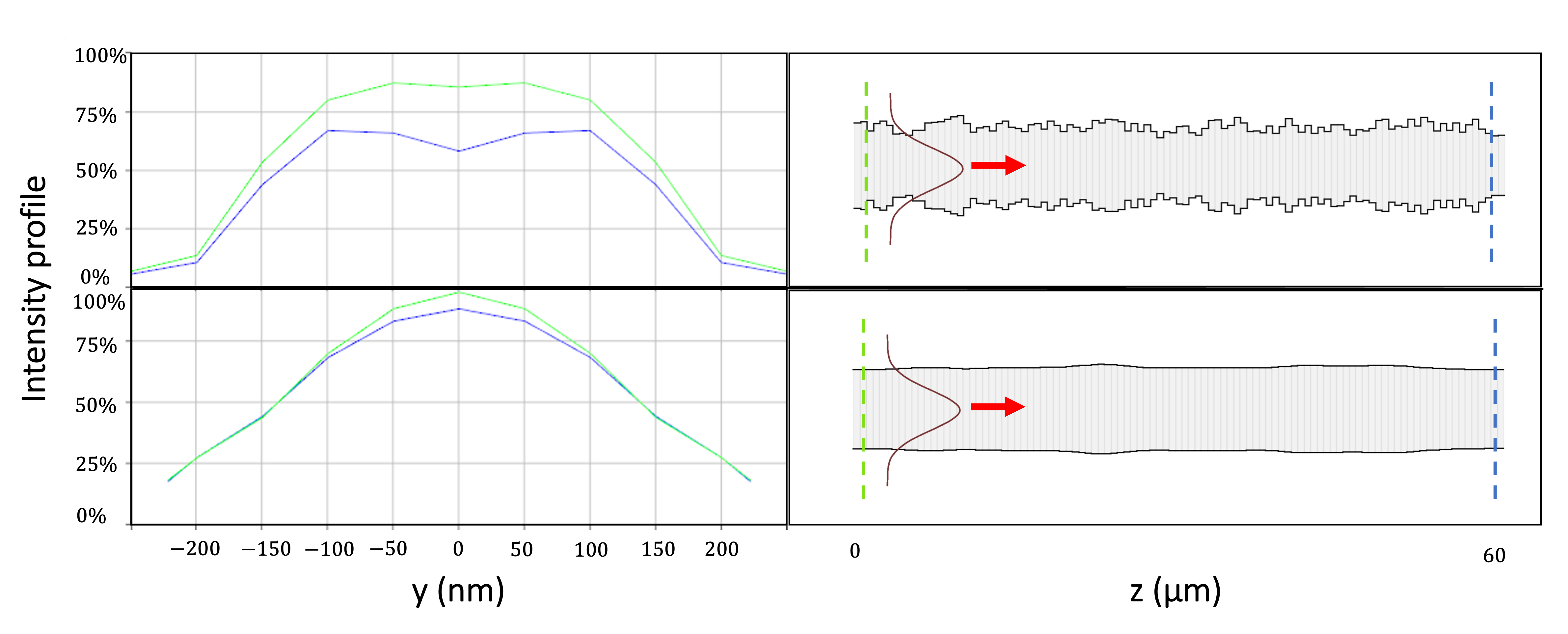}
    \caption{An illustration comparing baseline (top) and Fresnel-based (bottom) optimizations. On the left, the power density transmission cross-section is depicted using the FDTD Lumerical simulator, while on the right, the geometrical configuration along the propagation axis is shown. The green line corresponds to the intensity profile at the waveguide's entry, whereas the blue line indicates the final intensity profile. This configuration consists of 100 segments, using Silicon and a $\sqrt{X}$ gate, with a width noise standard deviation of $\sigma=30_{nm}$ ($\sigma_\% \approx 5\%$). Both designs achieve a fidelity of $0.91$.}
    \label{fig:power_dist_prof}
\end{figure*}

The averaged total power loss scaling is asymptotically linear in $N$:
\begin{gather}
    \lim_{N \rightarrow \infty} \langle P_{t} \rangle = \lim_{N \rightarrow \infty} \left(1- \left( 1-\langle R_{avg} \rangle \right)^{N-1}\right) 
    = N R_{avg} \ ,
\end{gather}
where $R_{avg}$ is the average reflection coefficient of a transition between two adjacent segments.
We plot in Fig. \ref{fig:power_dist_prof} the power density transmission cross-section curve using FDTD Lumerical simulator.

\subsection{Fidelity and Error Correlation}
\label{subsec:fidelity_and_error_correlation}

\begin{figure}
    \centering
    \includegraphics[width=1.0\linewidth]{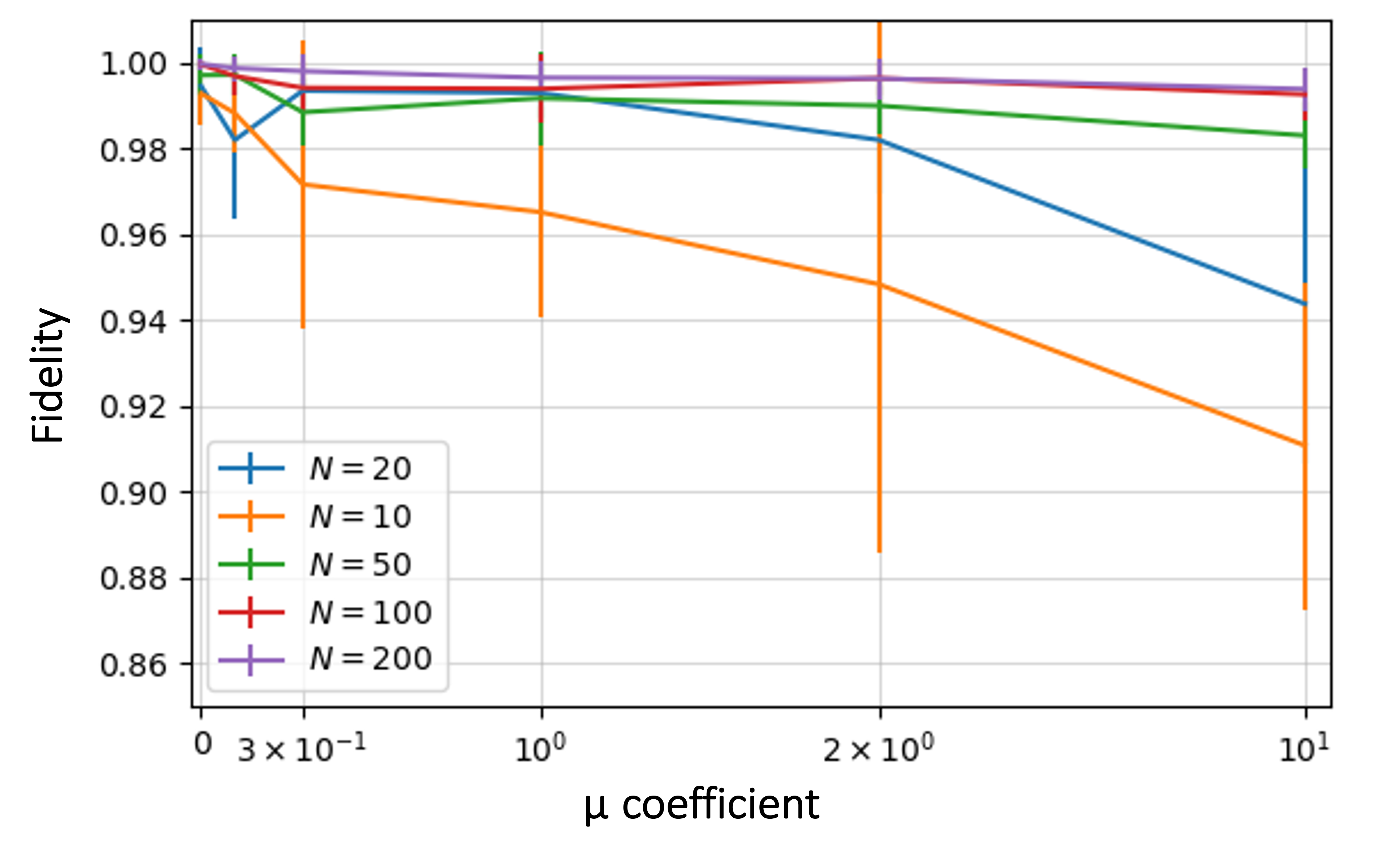}
    \caption{Mean value and standard deviation of the fidelity versus the error correlation strength (left - full correlation and right - low correlation), for different number of segments, for gate X, Silicon. We see the improvement of the design performance when increasing the correlation and the number of segments. This plot was created using an ensemble average over 10 optimizations for each $N$ and optimized for $\sigma_\% \approx 5\%$.}
    \label{fig:FvCorr}
\end{figure}

When the correlation between the errors at different segments decreases, the composite pulse scheme for error mitigation becomes less effective \cite{YaronIdoNoiseThreshold}. Here we analyze the impact of having a large number of segments when the error correlation decreases, where we model the error covariance matrix as Eq. \ref{rho}.
We plot in Fig. \ref{fig:FvCorr} the mean value and the standard deviation of the fidelity versus the error correlation strength $\mu$ (left, full correlation, and right, no correlation), for different numbers of segments for the Pauli X gate, Silicon material. As expected,  designs with a large number of segments show better performance
when the error correlations are small. We observe the same behavior for other quantum gates.

\subsection{Fidelity Versus Power Loss, and An Effective Error Mitigation Region}
\begin{figure}
    \centering
    \includegraphics[width=1.0\linewidth]{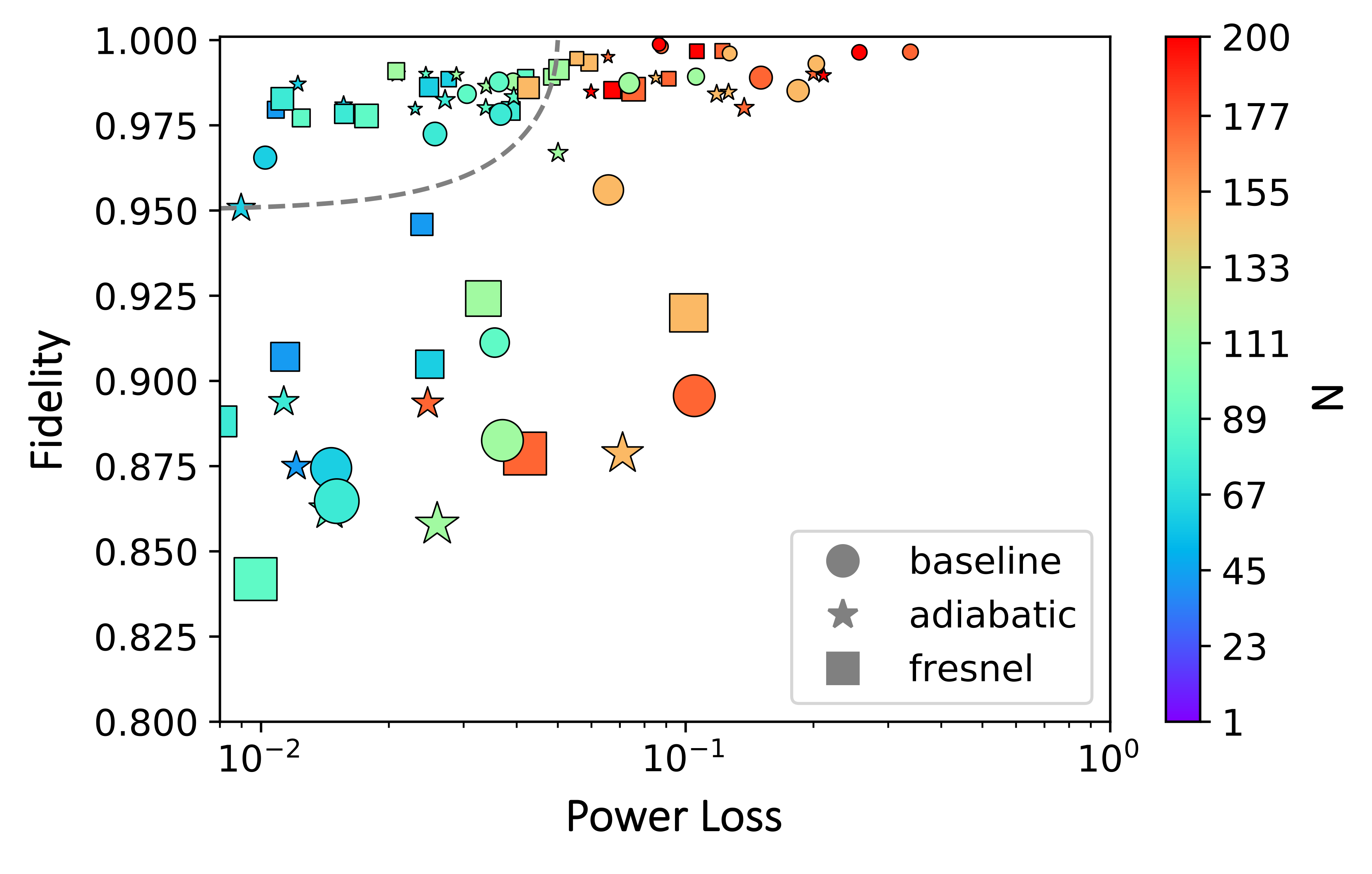}
    \caption{Fidelity versus power loss (As calculated with the FDTD simulation). Each point represents a single design, the color indicates the number of segments $N$, the shape indicates the optimization type, and the size of the point indicates the fidelity's standard deviation. The dashed line defines a domain that includes designs, where the mean fidelity is very high and the power loss is very low. Table \ref{tab:result_table} in appendix \ref{app:table_of_designs} gives the details of the designs.}
    \label{fig:PFdiagram}
\end{figure}

Taking into account both the fidelity of the quantum gate and the power loss of the design, we
plot in Fig. \ref{fig:PFdiagram} the mean fidelity versus the power loss. As seen, fidelity improves with a higher number of segments. While power loss also rises with more segments, it can be mitigated by incorporating the proposed smooth-design regularization.
The dashed line outlines a region we term as the \textit{Effective Error Mitigation Region}, as it encompasses exclusively designs that achieve both high fidelity and minimal power loss. All the designs presented in this figure are listed in table \ref{tab:result_table} in appendix \ref{app:table_of_designs}.

\section{Summary and Outlook}

We studied the performance of the segmented composite pulse error mitigation scheme applied to the design of photonic quantum gates, when increasing the number of segments $N$ and varying the error correlation strength $\mu$.
We quantified the impact of increasing $N$ on the fidelity and the power loss of the quantum gates and observed certain asymptotic scaling laws.
Empirically we found that $O(10)$ number of segments yields effective designs.

To minimize power loss with increasing $N$, we applied two optimization regularization: one imposing constraints on the Fresnel reflection coefficient and the other introducing an adiabatic condition on the waveguide width profile. Both approaches yielded comparable gate performance, although the adiabatic condition proves to be slightly more convenient within the optimization context.

We also examined the impact of varying correlation strength between errors in different segments. Our results showed that stronger correlations improve gate fidelity, as the segments can better compensate for accumulated errors. Notably, a large number of segments mitigates the challenges associated with weak error correlations.

There are several natural research directions to follow.
First, while our results are specifically tailored to dual-rail photonic gates, the insights gained from this study are expected to generalize across various noise models and error frameworks, including those with different error correlations. These findings hold potential for application to other quantum systems and platforms, providing valuable guidance for error mitigation strategies in broader quantum contexts.

Another important direction to follow is to study robust designs for QIP systems that include several or many quantum gates. In such cases, the optimization framework is likely to find robust designs that are better than a simple combination of individually robust single-qubit gates.

\vspace{0.5cm}
\textbf{Acknowledgements}
We would like to thank J. Drori and Y. Piasetzky for their assistance in extracting parameters using Lumerical simulations.
This work is supported in part by the Israeli Science Foundation Excellence Center No. 2312/21, the US-Israel Binational Science Foundation, and the Israel Ministry of Science.

\small
\bibliographystyle{unsrt.bst}
\bibliography{main}

\appendix

\section{Bloch Sphere Trajectories}
\label{app:bloch_sphere_trajectories}
\begin{figure}
    \centering
    \includegraphics[width=1.0\linewidth]{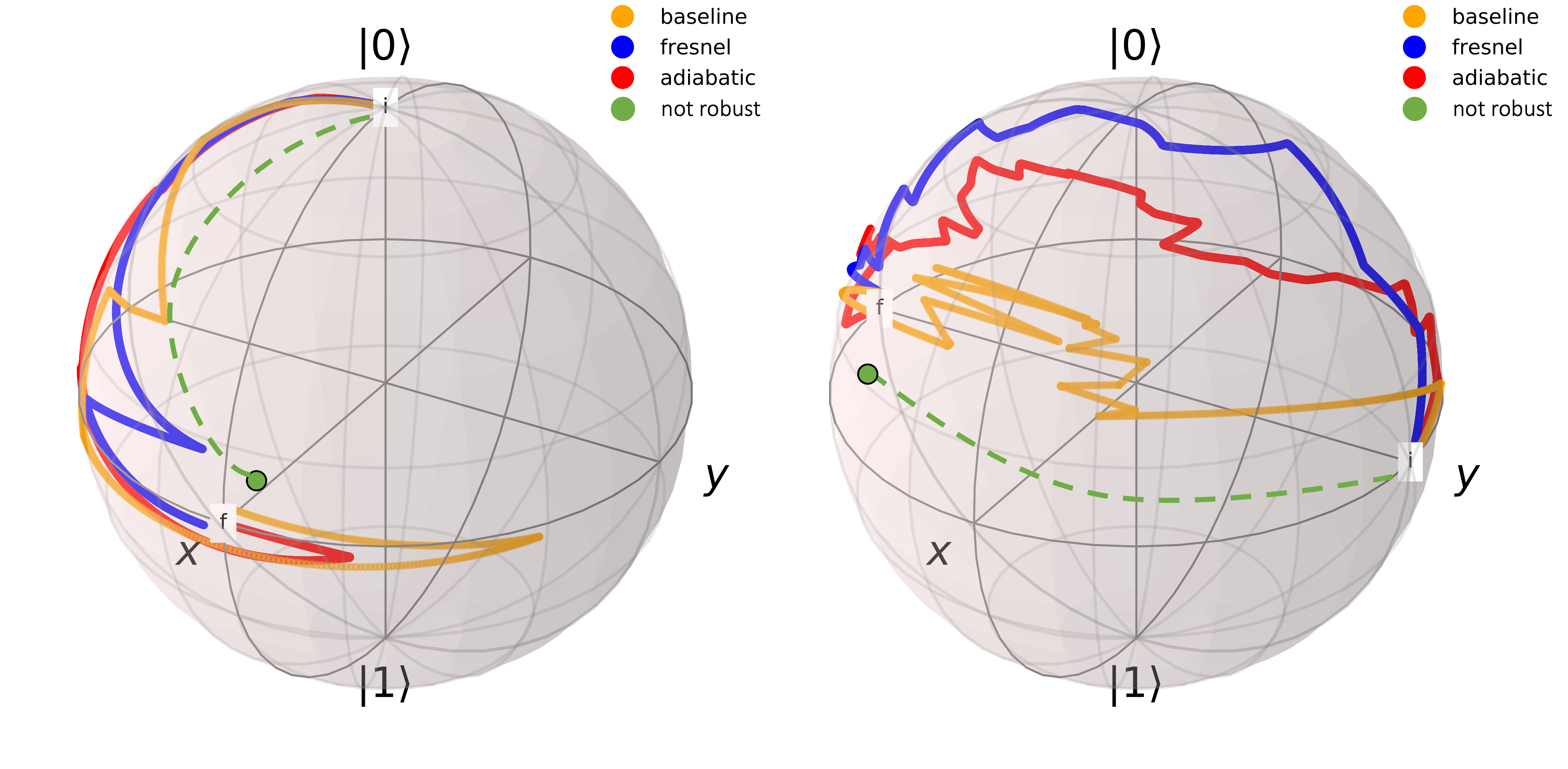}
    \caption{Bloch Sphere diagram of two configurations in Silicon. Left - Hadamard gate with 10 segments, initial state is $|0\rangle$ and final state is $|+\rangle$. Right - Pauli X gate with 50 segments, initial state is $|+\rangle_y$ and final state is $|-\rangle_y$.}
    \label{fig:BSphere}
\end{figure}

Fig. \ref{fig:BSphere} shows a trajectory comparison for two different cases.

\section{Power Loss Simulation Comparison}
\label{app:power_loss_sim_comp}
\begin{figure}
    \centering
    \includegraphics[width=1.0\linewidth]{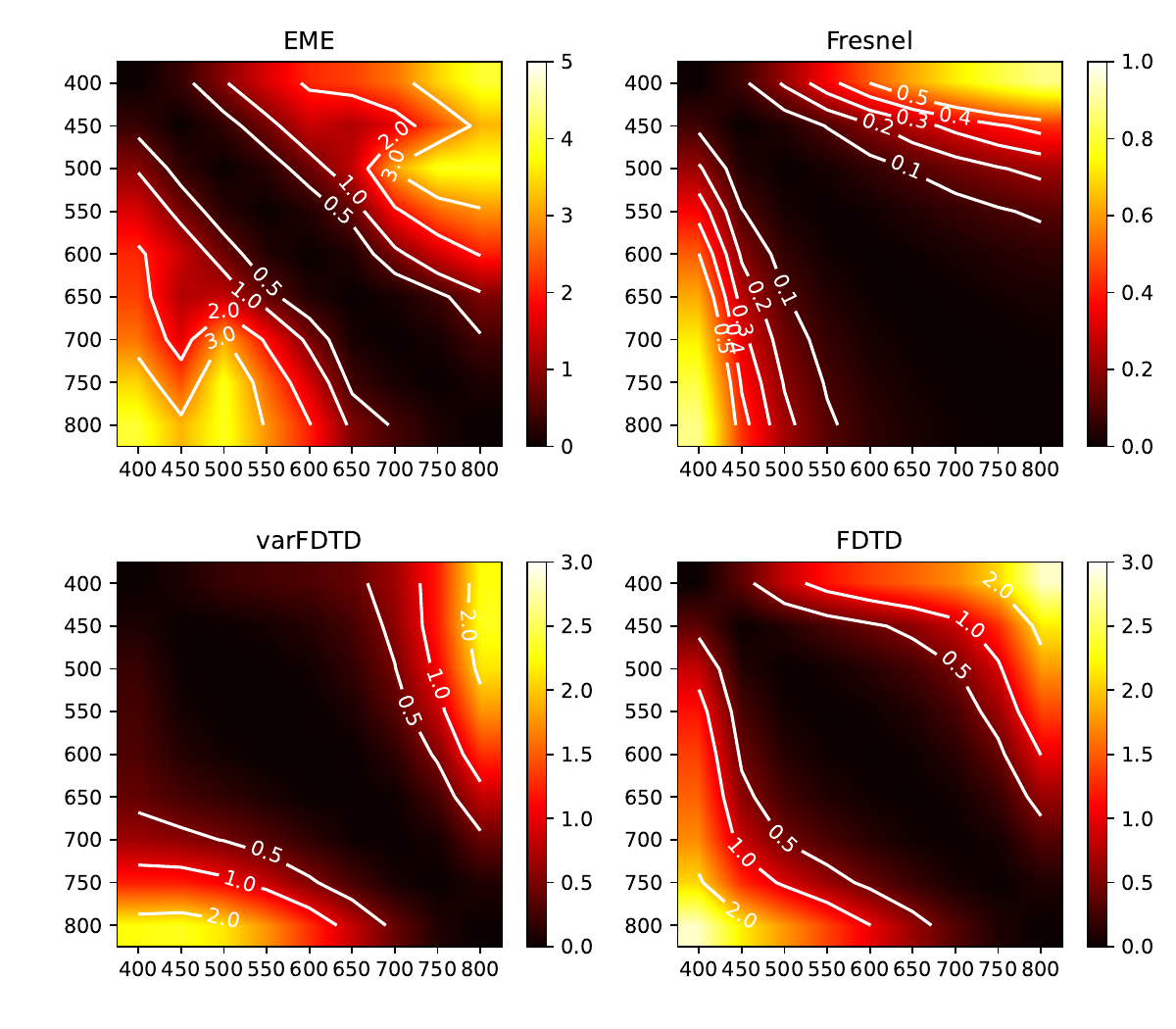}
    \caption{Our power loss simulation using EME (top-left), Fresnel coefficients (top-right), varFDTD (bottom-left) and FDTD (bottom-right). The simulation evaluated the power loss due to a single abrupt transition from width $W_0$ (x axis in nm) to width $W_1$ (y axis in nm), for Silicon and $H=220nm$. }
    \label{fig:power_loss_eme_fresnel}
\end{figure}

In Fig. \ref{fig:power_loss_eme_fresnel} we show a comparison between the power loss simulator (left) to the Fresnel coefficient (right). We consider the FDTD the most accurate simulation method and this is the reason we present this method along with the Fresnel coefficients. The Fresnel coefficient approximation of the power loss compared to the simulation, are related by a global factor of about 5 (see Fig. \ref{fig:PvsN} in the main text).

\section{Table of designs}
\label{app:table_of_designs}
In table \ref{tab:result_table} we present a list of designs of the Hadamard gate in Lithium Niobate and their fidelity and power loss.

\begin{table}
\centering
\caption{Table of Hadamard gate designs in Lithium Niobate. Mat stands for material and PL for Power loss.}
\label{tab:result_table}
\vspace{0.5cm}
\begin{tabular}{|l|l|l|l|l|l|l|l|l|}
\hline
Gate & Mat & Method & E[F] & STD[F] & Length & N & PL [$\%$] \\ \hline
Had & Li & adiabatic & 0.8466 & 0.0911 & 29$\mu m$ & 1 & 0.0000 \\ \hline
Had & Li & adiabatic & 0.8990 & 0.0775 & 39$\mu m$ & 3 & 0.7282 \\ \hline
Had & Li & adiabatic & 0.9486 & 0.1012 & 60$\mu m$ & 5 & 3.0097 \\ \hline
Had & Li & adiabatic & 0.9743 & 0.0747 & 81$\mu m$ & 7 & 3.7159 \\ \hline
Had & Li & adiabatic & 0.9727 & 0.0512 & 77$\mu m$ & 10 & 4.8779 \\ \hline
Had & Li & adiabatic & 0.9237 & 0.0419 & 45$\mu m$ & 20 & 9.9957 \\ \hline
Had & Li & adiabatic & 0.9853 & 0.0250 & 71$\mu m$ & 50 & 26.0394 \\ \hline
Had & Li & adiabatic & 0.9815 & 0.0329 & 78$\mu m$ & 100 & 27.0626 \\ \hline
Had & Si & adiabatic & 0.8709 & 0.1053 & 47$\mu m$ & 3 & 2.9732 \\ \hline
Had & Si & adiabatic & 0.9173 & 0.1081 & 51$\mu m$ & 5 & 1.5872 \\ \hline
Had & Si & adiabatic & 0.8908 & 0.1171 & 32$\mu m$ & 7 & 3.0901 \\ \hline
Had & Si & adiabatic & 0.8661 & 0.2202 & 45$\mu m$ & 10 & 3.7749 \\ \hline
Had & Si & adiabatic & 0.8699 & 0.2473 & 34$\mu m$ & 20 & 5.9694 \\ \hline
Had & Si & adiabatic & 0.8941 & 0.2066 & 141$\mu m$ & 50 & 15.2414 \\ \hline
Had & Si & adiabatic & 0.8878 & 0.1263 & 33$\mu m$ & 100 & 6.3382 \\ \hline
Had & Li & fresnel & 0.8813 & 0.0766 & 35$\mu m$ & 1 & 0.0000 \\ \hline
Had & Li & fresnel & 0.9044 & 0.0866 & 44$\mu m$ & 3 & 2.1209 \\ \hline
Had & Li & fresnel & 0.9812 & 0.0366 & 71$\mu m$ & 5 & 6.1240 \\ \hline
Had & Li & fresnel & 0.9748 & 0.0618 & 81$\mu m$ & 7 & 5.8759 \\ \hline
Had & Li & fresnel & 0.9803 & 0.0486 & 72$\mu m$ & 10 & 3.9069 \\ \hline
Had & Li & fresnel & 0.9899 & 0.0359 & 81$\mu m$ & 20 & 15.5735 \\ \hline
Had & Li & fresnel & 0.9863 & 0.0414 & 59$\mu m$ & 50 & 11.6337 \\ \hline
Had & Li & fresnel & 0.9873 & 0.0441 & 68$\mu m$ & 100 & 17.4665 \\ \hline
Had & Si & fresnel & 0.8170 & 0.1639 & 27$\mu m$ & 1 & 0.0000 \\ \hline
Had & Si & fresnel & 0.8908 & 0.1565 & 42$\mu m$ & 3 & 2.7992 \\ \hline
Had & Si & fresnel & 0.9028 & 0.0905 & 48$\mu m$ & 5 & 4.2536 \\ \hline
Had & Si & fresnel & 0.8942 & 0.1031 & 37$\mu m$ & 7 & 2.2589 \\ \hline
Had & Si & fresnel & 0.7502 & 0.2889 & 44$\mu m$ & 10 & 6.0034 \\ \hline
Had & Si & fresnel & 0.9161 & 0.1656 & 44$\mu m$ & 20 & 5.9888 \\ \hline
Had & Si & fresnel & 0.9261 & 0.1644 & 130$\mu m$ & 50 & 20.4289 \\ \hline
Had & Si & fresnel & 0.8844 & 0.2194 & 41$\mu m$ & 100 & 9.6211 \\ \hline
Had & Li & baseline & 0.9623 & 0.0801 & 60$\mu m$ & 1 & 0.0000 \\ \hline
Had & Li & baseline & 0.9713 & 0.0432 & 88$\mu m$ & 3 & 3.3101 \\ \hline
Had & Li & baseline & 0.9319 & 0.0321 & 58$\mu m$ & 5 & 6.8709 \\ \hline
Had & Li & baseline & 0.9768 & 0.0447 & 68$\mu m$ & 7 & 8.9922 \\ \hline
Had & Li & baseline & 0.9854 & 0.0370 & 81$\mu m$ & 10 & 11.1633 \\ \hline
Had & Li & baseline & 0.9727 & 0.0920 & 68$\mu m$ & 20 & 19.2621 \\ \hline
Had & Li & baseline & 0.9815 & 0.0567 & 66$\mu m$ & 50 & 38.9191 \\ \hline
Had & Li & baseline & 0.9899 & 0.0489 & 76$\mu m$ & 100 & 35.6861 \\ \hline
Had & Si & baseline & 0.8779 & 0.1106 & 45$\mu m$ & 1 & 0.0000 \\ \hline
Had & Si & baseline & 0.8571 & 0.2156 & 30$\mu m$ & 5 & 2.3238 \\ \hline
Had & Si & baseline & 0.8653 & 0.2521 & 41$\mu m$ & 7 & 3.6235 \\ \hline
Had & Si & baseline & 0.8957 & 0.2226 & 44$\mu m$ & 10 & 4.4218 \\ \hline
Had & Si & baseline & 0.8750 & 0.2121 & 69$\mu m$ & 20 & 6.9527 \\ \hline
Had & Si & baseline & 0.9517 & 0.1083 & 150$\mu m$ & 50 & 12.5530 \\ \hline
Had & Si & baseline & 0.8942 & 0.2124 & 112$\mu m$ & 100 & 22.0704 \\ \hline
\end{tabular}
\end{table}

\end{document}